\documentclass[journal]{IEEEtran}
\usepackage{cite}
\usepackage{graphicx}
\usepackage{subfigure}
\usepackage{url}
\usepackage{booktabs}
\usepackage{epsfig}
\usepackage{amsmath}
\usepackage{amsfonts}
\usepackage{amssymb}
\usepackage{mathrsfs}
\usepackage[final,scrtime]{prelim2e}  
\usepackage{verbatim}
\usepackage{algorithm}
\usepackage{algorithmic}
\usepackage{ntheorem}
\usepackage{url}
\usepackage{color}
\usepackage{multirow}     
\usepackage{diagbox}    
\usepackage{array}      
\usepackage{caption}    
\usepackage{colortbl}
\usepackage{longtable}   
\usepackage{xcolor}
\usepackage{xpatch}
\usepackage{lipsum,mwe,cuted}
\usepackage{stfloats}
\usepackage{enumerate}  
\usepackage{amsmath,mathtools}
\usepackage{makecell}

\graphicspath{{./}}
\theorembodyfont{\upshape}
\theoremheaderfont{\rmfamily\itshape}
\theoremseparator{:}

\theoremstyle{remark}

\newcommand{\tabincell}[2]{\begin{tabular}{@{}#1@{}}#2\end{tabula}}

\theoremstyle{remark}

\makeatletter
\def\changeBibColor#1{%
\in@{#1}{}
\ifin@\color{red}\else\normalcolor\fi
}
\xpatchcmd\@bibitem
{\item}
{\changeBibColor{#1}\item}
{}{\fail}
\xpatchcmd\@lbibitem
{\item}
{\changeBibColor{#2}\item}
{}{\fail}
\makeatother
\definecolor{r}{rgb}{1,0,0}
\definecolor{b}{rgb}{0,0,1}
\definecolor{k}{rgb}{0,1,1}

\def\mb{\mathbf}

\renewcommand{\vec}{\mb}    

\begin{document}
\title{Ray Antenna Array: A Novel Cost-Effective Multi-Antenna Architecture for Enhanced Wireless Communication}
\author{
	\IEEEauthorblockN{
		Zhenjun Dong\IEEEauthorrefmark{1},
		Zhiwen Zhou\IEEEauthorrefmark{1},
		Yong Zeng\IEEEauthorrefmark{1}\IEEEauthorrefmark{2},\\
	\IEEEauthorblockA{\IEEEauthorrefmark{1}National Mobile Communications Research Laboratory, Southeast University, Nanjing 210096, China\\}
	\IEEEauthorblockA{\IEEEauthorrefmark{2}Purple Mountain Laboratories, Nanjing 211111, China}
	\\Email: \{zhenjun\_dong, zhiwen\_zhou, yong\_zeng\}@seu.edu.cn}
}
\maketitle
\begin{abstract}
This paper proposes a novel multi-antenna architecture, termed ray antenna array (RAA), which aims to enhance wireless communication performance in a cost-effective manner.
RAA is composed of massive cheap antenna elements and a few radio frequency (RF) chains. The massive antenna elements are arranged in a novel ray-like structure, with each ray corresponding to a \emph{simple uniform linear array} (sULA) with a carefully designed orientation.
The antenna elements of each sULA are directly connected to an RF combiner, so that the sULA in each ray is able to form a beam towards a direction matching the ray orientation without relying on any analog or digital beamforming.
By further designing a ray selection network (RSN), appropriate sULAs are selected to connect to the RF chains for further baseband processing. Compared to conventional multi-antenna architectures like hybrid analog/digital beamforming (HBF), the proposed RAA has two major advantages.
First, it can significantly reduce hardware cost since no phase shifters, which are usually expensive especially in high-frequency systems, are required.
Besides, RAA can greatly improve system performance by configuring antenna elements with higher directionality, as each sULA only needs to be responsible for a portion of the total coverage angle.
To demonstrate such advantages, in this paper, we first present the input-output model for RAA-based wireless communications, based on which the ray orientations of the RAA are designed.
Furthermore, efficient algorithms for joint ray selection and beamforming are proposed for single-user and multi-user RAA-based wireless communications.
Simulation results demonstrate the superior performance of RAA  compared to HBF while significantly reducing hardware cost.
\end{abstract}

\section{Introduction}
The continuous advancement of multi-antenna or multiple-input multiple-output (MIMO) technology serves as a cornerstone in the evolution of wireless communication systems.
To harness the spatial degrees of freedom, the scale of antenna array is continuously expanding, enabling significant performance enhancements.
The forthcoming sixth-generation (6G) mobile communication networks are expected to  provide high-performance communications, environment sensing, and localization \cite{xiao2022overview}.
This renders the further advancement of multi-antenna technology critical, with various key technologies being investigated, such as extremely large-scale MIMO (XL-MIMO) \cite{10496996}, cell-free massive MIMO \cite{ngo2017cell}, and sparse massive MIMO \cite{li2024sparse}.
Despite the different evolution paths of multi-antenna technologies, a general consensus is that 6G will employ larger antenna arrays, higher frequency bands, and wider spectrum.
However, simply increasing the system dimension in spatial and time/frequency domains would inevitably conflict with other 6G objectives for improving energy/cost efficiency.
Therefore, developing advanced multi-antenna technology in a cost-effective manner, yet without degrading or even enhancing the performance, is of paramount importance.

As radio frequency (RF) chains and frontends are considered the most costly and power-hungry modules, an effective approach to reducing hardware and energy costs is to decrease the number of RF chains or use more cost-efficient RF components \cite{zeng2024tutorial}.
To this end, analog beamforming \cite{wang2009beam} and hybrid analog/digital beamforming (HBF) \cite{el2014spatially} have been extensively investigated, where one or a few RF chains are connected to massive antenna elements via phase shifter networks.
However, for 6G networks with large-scale arrays, the cost incurred by the phase shifters is substantial.
In particular, for high-frequency systems such as millimeter wave (mmWave) and Terahertz (THz) systems, designing accurate phase shifters is quite challenging.
To address such issues, antenna subset selection has been proposed \cite{7370753}, which replaces the expensive phase shifters with switches, but at the cost of performance degradation.
In \cite{zeng2016millimeter}, we proposed an electromagnetic (EM) lens-enabled array, which can focus the power of an incident wave on a small area of the antenna array, leading to a reduction in hardware and energy costs. 
Recently, several novel antenna architectures with reconfigurable antenna elements have been proposed, including fluid antenna (FA) \cite{wong2020fluid}, movable antenna (MA) \cite{zhu2023movable,dong2024movable}, and grouped MA \cite{lu2024group}.
These architectures offer the potential to significantly reduce the number of antenna elements and associated RF components.
However, they require additional hardware costs, such as motors, to control the movement of the antenna elements, leading to extra energy consumption and longer response time. 

In this paper, to further enhance performance while reducing hardware cost, we propose a novel multi-antenna architecture, termed {\it ray antenna array} (RAA).
The proposed RAA architecture is motivated by the fact that for an extremely {\it simple uniform linear array} (sULA) with all elements directly connected, without any analog or digital beamforming, a beam can still be formed in the direction matching the physical orientation of the sULA.
Therefore, by deploying multiple such sULAs with carefully designed orientations, flexible beamforming towards the desired direction can be achieved.

As shown in Fig. 1, the proposed RAA consists of multiple sULAs, which are deliberately arranged in a ray-like structure.
We deliberately use the term sULAs to highlight the fact that all antenna elements therein are directly connected to an RF combiner, as shown in Fig. 2.
Additionally, a ray selection network (RSN) is introduced to dynamically select the appropriate sULAs to be connected to the RF chains for further baseband processing.
Compared to the conventional HBF, the proposed RAA enjoys two significant advantages.
First, as antenna elements are much cheaper than phase shifters, RAA can significantly reduce hardware cost without requiring a single phase shifter, though more antenna elements are needed.
Second, RAA can significantly improve system performance by configuring antenna elements with stronger directionality, as each sULA only needs to cover a portion of the angle range.
Note that while RAA requires massive antenna elements to form its ray-like structure, which might increase its overall size, this is mitigated in high-frequency systems, where shorter wavelengths make it feasible to pack massive elements into reasonably compact form factors.

\section{Proposed Ray antenna array}
\begin{figure}[htbp]
\centering
\includegraphics[width=0.7\linewidth]{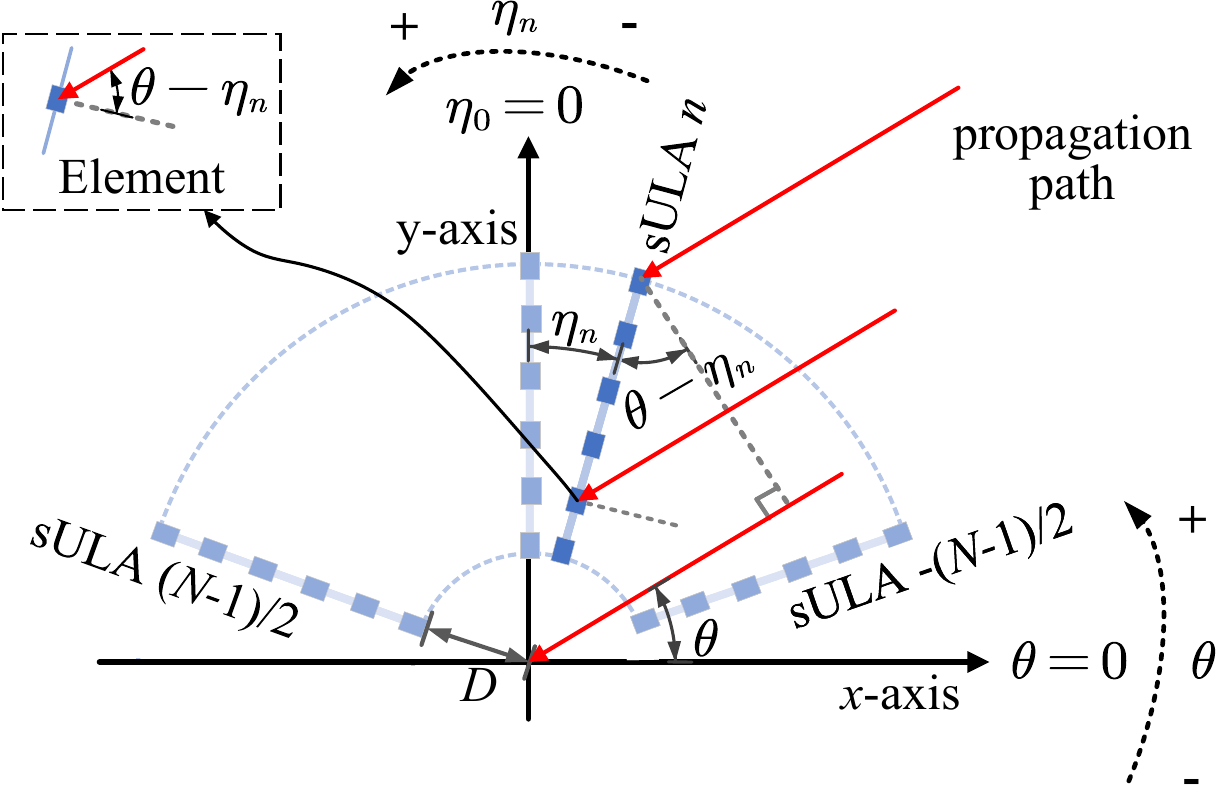}
	\caption{The proposed RAA consists of $MN$ cost-effective antenna elements, which are arranged into $N$ rays. Each ray corresponds to an $M$-element sULA, where all antenna elements are directly connected. }
	\label{array}
\end{figure}
\vspace{-0.2cm}
As illustrated in Fig. \ref{array}, the proposed RAA architecture consists of $NM$ antenna elements that are arranged into $N$ sULAs, where each sULA comprises $M$ elements with the adjacent elements separated by $d=\frac{\lambda}{2}$, with $\lambda$ being the signal wavelength.
The $N$ sULAs are arranged in a ray-like configuration, where each sULA has a carefully designed orientation.
Without loss of generality, we establish a Cartesian coordinate system such that the $N$ sULAs are placed symmetrically around the $y$-axis.
Let sULA $n=0$ align with the positive $y$-axis as the reference direction, and the orientation of the $n$th sULA w.r.t. the positive $y$-axis is denoted by $\eta_n\in\left[-\eta_{\text{max}},\eta_{\text{max}}\right]$, with $\eta_{\text{max}}$ denoting the maximum orientation and $\eta_{0}=0$, where $n\in\mathcal{N}$ with $\mathcal{N}=\left\{-\frac{N-1}{2},...,0,...,\frac{N-1}{2}\right\}$ being the set of sULA index.
For simplicity of notation, we assume $N$ is an odd number.
Note that $\eta_n$ is negative in the first quadrant and positive in the second quadrant.
To ensure a minimum separation for antenna elements across sULAs,
the first element of each sULA is positioned at a distance $D$ from the origin.
  Thus, the proposed RAA can be fully characterized by the parameters $\left(N,M,D,\{\eta_n\}_{n\in\mathcal{N}}\right)$.

 Denote by $\theta$ the angle of arrival/departure (AoA/AoD) of propagation path w.r.t the positive $x$-axis\footnote{Note that for notational convenience, the sULA orientation angle $\eta_n$ and the path angle $\theta$ are defined based on different reference angles.}, with $\theta\in\left[-\eta_{\text{max}},\eta_{\text{max}}\right]$.
Accordingly, the array response vector of the $n$th sULA for path angle $\theta$ can be expressed as
\begin{equation}\label{steering}
\vec{a}(\theta,\eta_n)=\bigl[1,e^{j\pi\sin(\theta-\eta_n)},...,e^{j\pi(M-1)\sin(\theta-\eta_n)}\bigl]^T,
\end{equation}
where $\theta-\eta_n$ is the AoA of the path relative to the $n$th sULA, as shown in Fig. \ref{array}. Accordingly, the array response matrix of the RAA, denoted by $\vec{A}(\theta)\in\mathbb{C}^{M\times N}$, is given by
\begin{equation}\label{steering_A}
\vec{A}(\theta)=\left[\vec{a}(\theta,\eta_n)\right]_{n\in\mathcal{N}}\times\text{diag}\left(\vec{b}\right),
\end{equation}
where $\vec{b}=\left[b(\theta,\eta_n)\right]_{n\in\mathcal{N}}\in\mathbb{C}^{N\times 1}$ captures the response of the reference antenna element of each sULA, e.g., the first element, given by
\begin{small}
\begin{equation}\label{steering_bn}
b(\theta,\eta_n)=e^{j\frac{2\pi}{\lambda}D\sin(\theta-\eta_n)}\sqrt{G(\theta-\eta_n)}.
\end{equation}
\end{small}%

Note that in (\ref{steering_bn}), the first term represents the phase shift of
the first element w.r.t. the origin, while the second term $G(\theta-\eta_n)$ is the radiation pattern of each antenna element with $G(0)$ corresponding to the peak point, which is determined by the 3dB bandwidth and the maximum radiation power. 
From (\ref{steering}) and (\ref{steering_A}), it can be observed that by selecting the appropriate sULAs to connect to the RF chains, a flexible beamforming towards the desired direction can be achieved without relying on any analog or digital beamforming.

\vspace{-0.2cm}
\begin{figure}[htbp]
\centering
	\includegraphics[width=0.7\linewidth]{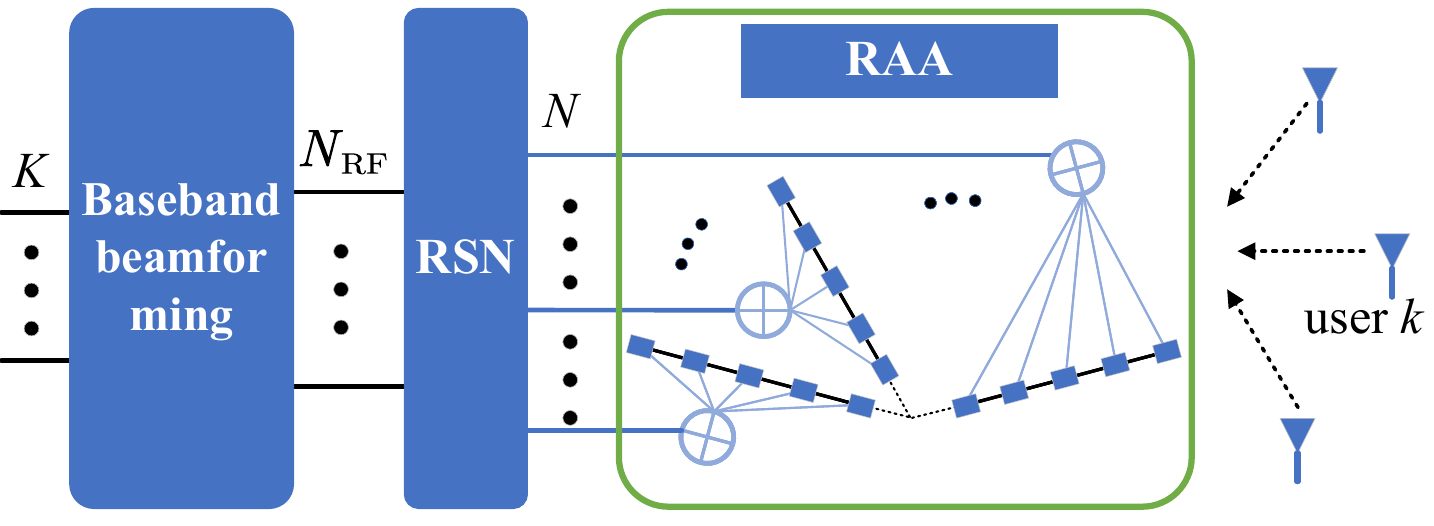}
\caption{RAA-based wireless communications.}
\label{architecture3X}
\end{figure}
\vspace{-0.2cm}

A wireless communication system based on the proposed RAA with $NM$ elements and $N_{\text{RF}}$ RF chains is illustrated in Fig. \ref{architecture3X}, with $N_{\text{RF}} \leq N$. All $M$ elements of each sULA are connected to an RF combiner.
The resulting outputs for all $N$ sULAs, denoted by $\vec{r}(\theta)=\left[r(\theta,\eta_n)\right]_{n\in\mathcal{N}}\in\mathbb{C}^{N\times 1}$, are expressed as
\begin{equation}\label{steering_eff00}
\vec{r}(\theta)=\vec{A}^T(\theta)\vec{1}_{M\times 1}=\text{diag}\left(\vec{b}\right)\bigl[H_M(\sin(\theta-\eta_n)) \bigl]_{n\in\mathcal{N}},
\end{equation}
where $H_M\left(\sin(\theta-\eta_n)\right)=\vec{a}^T(\theta,\eta_n)\vec{1}_{M\times 1}$ is the Dirichlet kernel function given by
\begin{small}
\begin{equation}\label{H_M}
H_M\left(\sin(\theta-\eta_n)\right)=e^{j\frac{\pi}{2}(M-1)\sin(\theta-\eta_n)}\frac{\sin\left(\frac{\pi}{2}M\sin(\theta-\eta_n)\right)}{\sin\left(\frac{\pi}{2}\sin(\theta-\eta_n)\right)}.
\end{equation}
\end{small}%

An RSN is designed to select $N_{\text{RF}}$ ports from the $N$ equivalent outputs. The RSN selection matrix
is denoted by $\vec{S}\in\{0,1\}^{N_{\text{RF}}\times N}$,
which satisfies $\left\|[\vec{S}]_{i,:}\right\|_0=1$ and $\left\|[\vec{S}]_{:,n}\right\|_0\leq1$
, $1\leq i\leq N_{\text{RF}}$ and $1\leq n\leq N$. Thus, the set of $N_{\text{RF}}$ selected elements' indices is denoted by $\Omega=\{n:[\vec{S}]_{i,n}=1, 1\leq i\leq N_{\text{RF}}\}$.


\section{RAA Orientation Design and Comparison with HBF}

In this section, we design the orientations $\{\eta_n\}_{n\in\mathcal{N}}$ of the RAA and compare the cost against HBF. First, to achieve the null points of $H_M\left(\sin(\theta-\eta_n)\right)$ in (\ref{steering_eff00}), it is required that
$\frac{\pi}{2} M\sin(\theta-\eta_n)=p \pi$ ($p=\pm1, \pm2,...$) in (\ref{H_M}), which yields the null points, i.e., $\theta-\eta_n=\arcsin\left(2p/M\right)$.
For sULA $n=0$, the peak point can be obtained at $\theta=0$ due to $\eta_n=0$, and
the null points can be obtained at $\theta=\arcsin\left(\frac{2p}{M}\right)$.
Thus, the peak-to-null beamwidth of the main lobe for sULA $n=0$ is $\arcsin\left(\frac{2}{M}\right)$.
However, due to the nonlinear nature of the function $\arcsin(\cdot)$, the null points are not evenly spaced, i.e., $\arcsin\left(\frac{2p}{M}\right)\neq p\times\arcsin\left(\frac{2}{M}\right)$.
As a result, the angle interval between adjacent null points of sULA $n=0$ is not a constant.
Consequently, it is impossible to perfectly align the null points of sULA $n=0$ with those of sULA $n$ for any $n\neq 0$.
Therefore, to minimize interference between adjacent sULAs, their orientation is designed to satisfy the criterion that
the null point of one sULA's main lobe aligns with the peak of the adjacent sULA's main lobe, i.e.,
\vspace{-0.05cm}
\begin{equation}
\eta_n=n\times\arcsin\left(2/M\right), \enspace \forall n\in\mathcal{N}.
\vspace{-0.05cm}
\end{equation}

Thus, the required number of sULAs in the proposed RAA is given by $N=2 \bigl\lfloor\frac{\eta_{\text{max}}}{\arcsin(2/M)}\bigl\rfloor+1$ with $\lfloor\cdot\rfloor$ denoting the floor function.
 In addition, to ensure that the distance between the first antennas of adjacent sULAs is greater than half a wavelength, the distance $D$ needs to satisfy
$D\geq \lambda/(4\sin(0.5\arcsin(2/M)))$.




Next, we compare the RAA with the DFT codebook-based HBF. 
The array response vector of an $M$-element ULA is given by $\vec{a}(\theta)=[e^{j\pi(m-1)\sin\theta}]_{1\leq m\leq M}\in \mathbb{C}^{M\times 1}$.
 The DFT codebook used in the HBF architecture is denoted by $\vec{A}_{\text{DFT}}=[\vec{a}_{\text{DFT}}(\theta_n)]_{1\leq n\leq N'}\in \mathbb{C}^{M\times N'}$, where $\vec{a}_{\text{DFT}}(\theta_n)=[e^{j\pi(m-1)\sin\theta_n}]_{1\leq m\leq M}\in \mathbb{C}^{M\times 1}$ is the $n$th DFT codeword and $N'$ is the number of DFT codewords.
 Consequently, considering the antenna pattern, the beam pattern for the HBF, denoted by $\vec{r}_{\text{HBF}}(\theta)=[r(\theta,\theta_n)]_{1\leq n\leq N'}$, can be expressed as
\begin{equation}\label{steeringV1X}
\vec{r}_{\text{HBF}}(\theta)
=\sqrt{G_{\text{ULA}}(\theta)}\left[H_M(\sin\theta-\sin\theta_n)\right]_{1\leq n\leq N'},
\end{equation}
where $G_{\text{ULA}}(\theta)$ is the radiation pattern of each antenna element in the $M$-element ULA under the HBF architecture. 

By comparing (\ref{steering_eff00}) with (\ref{steeringV1X}), it is observed that when the AoA $\theta$ is perpendicular to sULA $n$, i.e., $\theta=\eta_n$, the RAA can obtain $M$-times array gain similar to the HBF.
However, the RAA only requires low-cost antenna elements and RF combiners rather than expensive phase shifters.
Moreover, the RAA samples the AoA $\theta$ uniformly with an interval of $\arcsin\left(\frac{2}{M}\right)$, in contrast to  the HBF which instead samples $\sin(\theta)$ uniformly with an interval of $\frac{2}{M}$ \cite{pillai2012array}.
This uniform angle sampling enables the RAA to achieve enhanced spatial resolution and consistent coverage across the angular domain.

Compared to the fully-connected HBF architecture that requires $MN_{\text{RF}}$ phase shifters and $M$ antennas, the proposed RAA architecture requires $N_{\text{RF}}N$ switches and $NM$ antenna elements.  
Since phase shifters are typically much more expensive than antenna elements and switches, the RAA can significantly reduce the overall hardware cost. 
The major hardware costs of the RAA and HBF are given by
$\text{cost}_{\text{RAA}}=N_{\text{RF}}Np_{\text{sw}}+NMp_{\text{ant}}$ and $
\text{cost}_{\text{HBF}}=N_{\text{RF}}Mp_{\text{ps}}+Mp_{\text{ant}}$,
 respectively, where $p_{\text{sw}}$, $p_{\text{ant}}$, and $p_{\text{ps}}$ are the unit prices of RF switch, antenna element, and phase shifter, respectively. 
 Since 
 $p_{\text{ps}}$, $p_{\text{sw}}$, and $p_{\text{ant}}$ are typically on the order of $100 \$$, $0.1 \$$, and $0.01 \$$, respectively, it is not difficult to prove that the proposed RAA can reduce the hardware cost, i.e., $\text{cost}_{\text{RAA}}< \text{cost}_{\text{HBF}}$.
 As an example, we consider a sub-6 GHz communication system that employs the TGP2108-SM 6-bit digital phase shifter and the QM12002 RF switch, both manufactured by Qorvo \cite{qorvo2025}.
Based on the manufacturers' quotations, we have $p_{\text{ps}}\approx63.44 \$$, $p_{\text{sw}}\approx0.12 \$$, and $p_{\text{ant}} \approx 0.01 \$$.
In addition, $16$-element ULA and $\eta_{\text{max}}=0.5\pi$ are considered, thus we have $N=25$.
For a single RF chain, the hardware costs of the RAA and HBF are $7\$$ and $1015.2\$$, respectively.
Therefore,
the proposed RAA can significantly reduce the hardware cost, which accounts for only 6.9\textperthousand$ $ of that of the HBF architecture. 

\section{Joint RAA Beamforming and Ray Selection}

As shown in Fig. 2, the uplink RAA-based multi-user communications are considered, where a base station (BS) equipped with the proposed RAA serves $K$ single-antenna users.
The resulting signal of user $k$ at the BS is expressed as
\begin{equation}\label{channel model}
    y_k=\vec{f}_k^H \vec{S}\vec{h}_ks_k+\vec{f}_k^H \vec{S}\sum\nolimits_{i\neq k}\vec{h}_is_i
    +\vec{f}_k^H \vec{S}\vec{z}',
\end{equation}
where $\vec{f}_k\in \mathbb{C}^{N_{\text{RF}} \times 1}$ is the baseband beamforming of user $k$;
$\vec{h}_k\in\mathbb{C}^{N\times 1}$ is the effective channel vector for user $k$, modeled as
\begin{equation}\label{h_eff}
\vec{h}_k=\sum\nolimits_{l =1}^{L_k}\alpha_{k,l} \vec{r}(\theta_{k,l}),
\end{equation}
where $L_k$ is the number of multipath components (MPCs) for user $k$,
$\alpha_{k,l}$ is the complex gain of the $l$th MPC for user $k$, 
and
 $\theta_{k,l}\in \left[-\eta_{\text{max}},\eta_{\text{max}}\right]$ is the AoA of the $l$th MPC for user $k$;
 $s_k$ is the information-bearing symbol of user $k$ with $\mathbb{E}[|s_k|^2]=P_t$,  and $P_t$ is the transmit power;
 and $\vec{z}'=[\vec{z}_n^T\vec{1}_{M\times 1}]_{n\in\mathcal{N}}\in \mathbb{C}^{N\times 1}$ is the effective noise vector, with $\vec{z}_n\in \mathbb{C}^{M\times 1}\sim N_{\mathbb{C}}\left(0,\sigma^2\vec{I}_{M}\right)$ being the noise at sULA $n$ following independent and identically distributed (i.i.d.) circularly symmetric complex gaussian (CSCG) distribution with variance $\sigma^2$, thereby  $\vec{z}' \sim N_{\mathbb{C}}(0,M\sigma^2\vec{I}_{N})$.
Thus, the signal-to-interference plus noise ratio (SINR) of user $k$ is
\begin{equation}\label{channel model3}
     \text{SINR}_k=\frac{\bar{P_t}\left|\vec{f}_k^H \vec{S}\vec{h}_k
    \right|^2}{\bar{P_t}\sum_{i\neq k}\left|\vec{f}_k^H \vec{S}\vec{h}_i
    \right|^2+M\|\vec{f}_k^H \vec{S}\|^2},
\end{equation}
where $\bar{P_t}=\frac{P_t}{\sigma^2}$ represents the transmit signal-to-noise ratio (SNR).
In particular, the SNR for the single-user scenario is
 \begin{equation}\label{SNR}
\text{SNR}=\bar{P_t}\left|\vec{f}^H\vec{S}\vec{h}\right|^2/(M\vec{f}^H \vec{S}\vec{S}^H \vec{f})\triangleq\bar{P_t}\left|\vec{f}^H\vec{S}\vec{h}\right|^2/(M\vec{f}^H  \vec{f}),
 \end{equation}
where ``$\triangleq$" follows from $[\vec{S}]_{i,:}[\vec{S}]_{i,:}^H=1$ and $[\vec{S}]_{i,:}[\vec{S}]_{j,:}^H=0$, $\forall j\neq i$, leading to $\vec{S}\vec{S}^H=\vec{I}_{N_{\text{RF}}}$
. For any given ray selection matrix $\vec{S}$, the optimal baseband beamforming is the
maximum ratio combining (MRC), i.e., $\vec{f}=\vec{S}\vec{h}$.
Thus, the SNR can be expressed as a function of the ray selection matrix $\vec{S}$ as
 \begin{equation}\label{SNR2}
\text{SNR}(\vec{S})=\bar{P_t}\|\vec{S}\vec{h}\|^2/M.
 \end{equation}

 It is observed from (\ref{SNR2}) that to maximize the SNR in a single-user RAA system, the optimal $\vec{S}$ corresponds to selecting the $N_{\text{RF}}$ elements of $\mathbf h$ with the largest magnitudes.
In particular, considering a single-path case, we have $h_n=\alpha_l\sqrt{G(0)}M$ when the path is perpendicular to sULA $n$, i.e., $\theta=\eta_n$. Thus, the resulting maximum SNR in (\ref{SNR2}) is $\bar{P_t}|\alpha_l|^2G(0)M$, which increases with $M$. 
Similar conclusions can be drawn for the multipath case.

For the more general multi-user RAA system, the joint design of $\vec{F}=\left[\vec{f}_k\right]_{1\leq k \leq K}\in \mathbb{C}^{ N_{\text{RF}}\times K}$ and $\vec{S}$ is formulated as an optimization problem to maximize the sum rate
\vspace{-0.2cm}
\begin{equation}\label{channel model4}
\begin{aligned}
   \underset{\vec{F},\vec{S}}{\text{max}}& \quad R_{\text{sum}}=\sum\nolimits_{k=1}^{K}\text{log}_2(1+\text{SINR}_k)\\
   \text{s.t.}& \quad(C1):
               \enspace\vec{S}\in \{0,1\}^{N_{\text{RF}}\times N},\\
               &\quad(C2):\enspace\left\|[\vec{S}]_{i,:}\right\|_0=1,\enspace 1\leq i\leq N_{\text{RF}},\\
               &\quad(C3):\enspace\left\|[\vec{S}]_{:,n}\right\|_0\leq 1,\enspace 1\leq n\leq N.\\
   \end{aligned}
\end{equation}

Problem (\ref{channel model4}) is challenging to solve directly due to the binary constraint in (C1) and the non-convex $l_0$-norm in (C2) and (C3).
In addition, the digital beamforming $\vec{F}$ and ray selection $\vec{S}$ are coupled.
It is worth noting that if the exhaustive search is applied to select the $N_{\text{RF}}$ best rays from a total of $N$, it will result in $\tbinom{N}{N_{\text{RF}}}$ candidate solutions, leading to prohibitive complexity when $N$ and ${N_{\text{RF}}}$ are large.
Therefore, a greedy scheme is proposed to reduce the computational complexity to $O(NN_{\text{RF}})$.

Note that for any given ray selection matrix $\vec{S}$, it follows from (\ref{channel model3}) that the optimal baseband beamforming for sum rate maximization is given by the minimum mean square error (MMSE) beamforming, which is expressed as
$\vec{f}_k=\vec{C}_k^{-1}\vec{S}\vec{h}_k$ with $\vec{C}_k\triangleq\vec{S} \bigl(\sum_{i \neq k}\vec{h}_i\vec{h}_i^H+M/\bar{P_t}\vec{I}_N\bigl)\vec{S}^H$ being the interference-plus-noise covariance matrix, $1\leq k\leq K$.
By substituting the matrices $\vec{S}$ and $\vec{F}$ into (\ref{channel model4}), the resulting maximum sum rate under the set $\Omega$ is
\vspace{-0.05cm}
\begin{equation}\label{channel model5}
    R_{\text{sum}}(\Omega)=\sum\nolimits_{k=1 }^{K}\text{log}_2\bigl(1+(\vec{S}\vec{h}_k)^H \vec{C}_k^{-1}\vec{S}\vec{h}_k\bigl).
\end{equation}

In the proposed greedy scheme, one ray of the RAA is selected at each step to maximize the sum rate.
At the $i$-th step, the optimal element $n^{(i)}$ is determined based on $n^{(i)}=\underset{n}{\text{argmax}} \quad R_{\text{sum}}(\Omega^{(i-1)}\cup n)$,
and the resulting sum rate is $R_{\text{sum}}(\Omega^{(i)})$ with $\Omega^{(i)}=\Omega^{(i-1)}\cup n^{(i)}$. This process continues until $N_{\text{RF}}$ sULAs are selected.
The main procedure of the greedy scheme is summarized in Algorithm 1.
\vspace{-0.1cm}
\begin{algorithm}[htb]
\caption{The proposed greedy scheme for RAA ray selection}
\label{alg:Framwork}
\begin{algorithmic}[1] 
\STATE Initialize $\Omega^{(0)}=\emptyset$ and $\mathcal{N}=\{1,..,N\}$.\\ 
\STATE \textbf{for} $i=1:N_{\text{RF}}$ \textbf{do}\\ 
\STATE  \quad Calculate $n^{(i)}=\underset{n\in \mathcal{N}}{\text{argmax}} \quad R_{\text{sum}}(\Omega^{(i-1)}\cup n)$;
\STATE \quad Update $\Omega^{(i)}=\Omega^{(i-1)}\cup n$ and $\mathcal{N}=\mathcal{N}\backslash n^{(i)}$;
\STATE \textbf{end for}
\STATE \textbf{Output:} The set of selected $N_{\text{RF}}$ elements $\Omega^{(N_{\text{RF}})}$.
\end{algorithmic}
\end{algorithm}

\vspace{-0.5cm}
\section{Simulation Results}
In this section, simulation results are provided to demonstrate the performance of the proposed RAA.
Unless otherwise specified, the number of elements in each sULA is set to $M=16$, and the maximum orientation is set to $\eta_{\text{max}}=0.5\pi$.
Thus, we have $N=25$ and $\eta_n=n\times\arcsin\left(1/8\right)$, $\forall n\in\mathcal{N}$.
In addition, for the DFT codebook, we have $N'=\frac{2}{1/8}=16$ and $\theta_n=\arcsin\bigl(\frac{n}{8}\bigl)$, $-8\leq n\leq 7$.
The number of RF chains is set to $N_{\text{RF}}=5$, and the transmit SNR varies from -10 dB to 10 dB. In addition, the antenna radiation pattern adopts the 3GPP antenna model \cite{3GPPchannel}, which is expressed in dB as
\vspace{-0.05cm}
\begin{equation}\label{G}
G(\theta)=G_0^{\text{dB}}-\text{min}\bigl\{12\left(\theta/\theta_{\text{3dB}}\right)^2,A_{\text{max}}^{\text{dB}}\bigl\},
\vspace{-0.05cm}
\end{equation}
where $G_0^{\text{dB}}$ denotes the peak antenna gain in dB, $A_{\text{max}}^{\text{dB}}=$30 dB describes the front-to-back attenuation and $\theta_{\text{3dB}}$ accounts for the 3 dB beamwidth.
$\theta_{\text{3dB}}$ in the HBF is set to $\pi$ to cover the entire AoA range,
while in the RAA it is set to $0.3\pi$, as each sULA only
needs to be responsible for a narrower angle range.
Furthermore, $G_0^{\text{dB}}$ in the HBF is set to 0 dB, while in the RAA it is set to 5.1335 dB to ensure the same radiation power as that of the HBF.
In addition,
the isotropic antenna element gain in any direction is set to -2.816 dB to match the radiation power of the directional antenna.
The parameter settings for single- and multi-user scenarios are defined in Table II.
\vspace{-0.1cm}
\begin{table}[htbp]
\caption{The parameter settings.}
\centering
\begin{tabular}{|c|c|c|}
\hline
\multirow{2}{*}[-1.5ex]{\makecell[c]{Single\\user }} &  MPCs' number  & $L=5$      \\ \cline{2-3}
                  &     \makecell[c]{The $l$th MPC,\\$1\leq l\leq L$} & \makecell[c]{$\theta_l=-0.5\pi+(0.15\pi)l$, \\$|\alpha_l|=\sqrt{0.2}$, $\frac{\alpha_l}{|\alpha_l|}$: random} \\ \hline
\multirow{2}{*}[-4ex]{\makecell[c]{Multi\\ user}} &  Users' number  &  $K=5$ \\ \cline{2-3}
                  & \multicolumn{1}{l|}{\makecell[c]{User $k$,\\$1\leq k\leq K$}} & \multicolumn{1}{l|}{\makecell{$L_k=2$,  $|\alpha_{k,l}|=\sqrt{0.5}$, $\frac{\alpha_{k,l}}{|\alpha_{k,l}|}$: random
                  \\$\theta_{k,l}\sim N\left(\bar{\theta}_k,(0.1\pi)^2\right)$,\\ $\bar{\theta}_k=-0.5\pi+(0.15\pi)k$}} \\ \hline
\end{tabular}
\end{table}
\vspace{-0.1cm}

\begin{figure}[htbp]
\centering
\subfigure[]{
	\includegraphics[width=0.55\linewidth]{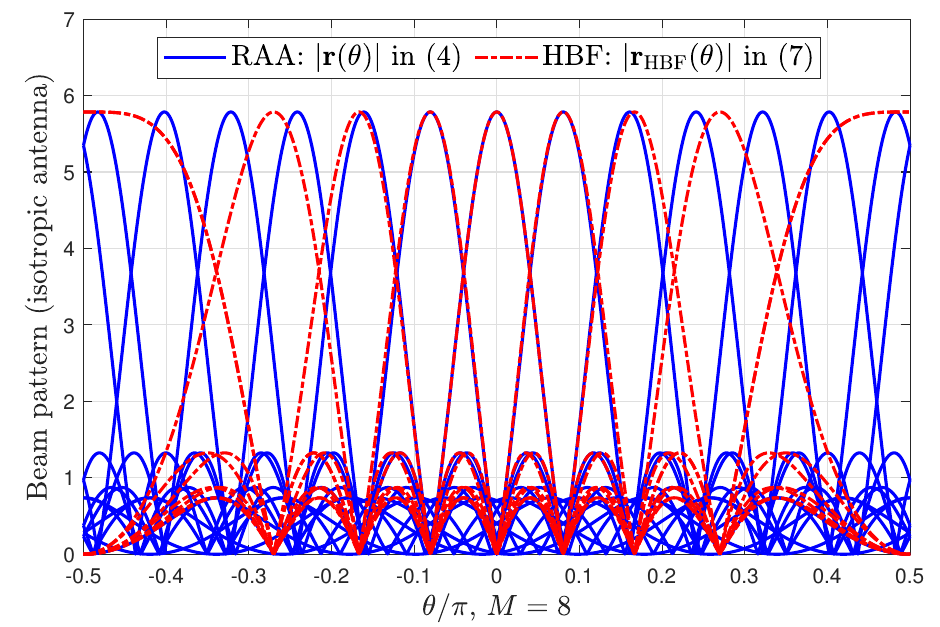}}
\subfigure[]{
	\includegraphics[width=0.55\linewidth]{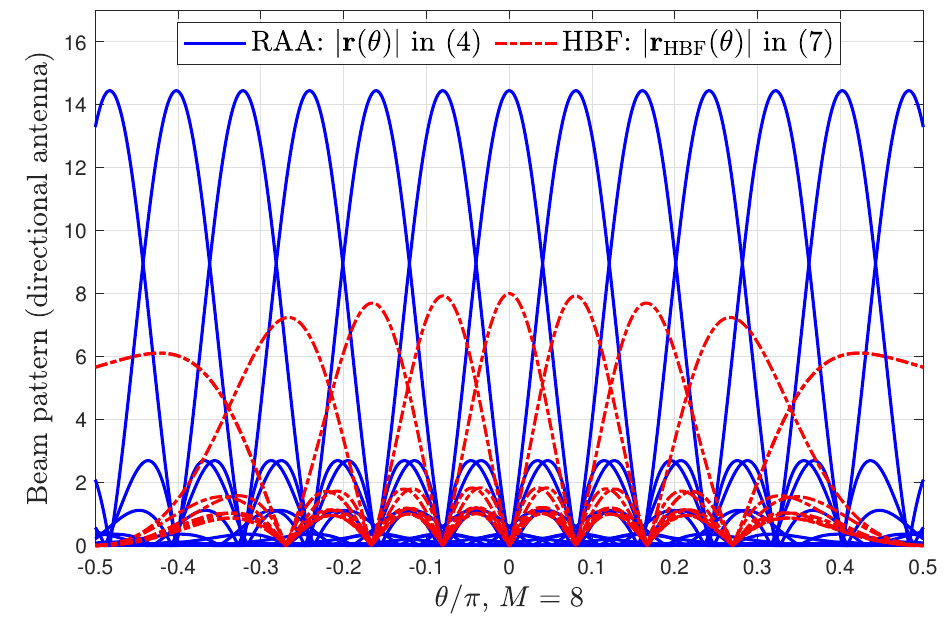}}
\caption{Beam patterns of the proposed RAA and HBF, considering (a) isotropic and (b) directional antenna element, where $M=8$ and $N=13$.}
\label{beam_pattern}
\vspace{-0.45cm}
\end{figure}

Fig. \ref{beam_pattern} presents the beam patterns of the proposed RAA and the HBF, considering (a) isotropic and (b) directional antenna elements, where $M=8$ and $N=13$.
From Fig. \ref{beam_pattern}(a), it can be observed that when $|\theta|<0.1\pi$, the proposed RAA has approximately the same beam pattern as the HBF.
This confirms that each sULA can
achieve efficient beamforming without relying on any analog or digital beamforming.
It is also observed from Fig. \ref{beam_pattern} that for $|\theta|>0.16\pi$, the main lobe's beamwidth for the RAA is narrower than that for the HBF.
This difference arises because the RAA uniformly samples spatial angle $\theta$, whereas the HBF samples $\sin(\theta)$ uniformly.
Furthermore, it is found from Fig. \ref{beam_pattern}(b) that the beam gain of the RAA is significantly higher than that of HBF.
This is attributed to the higher directional antenna element gain of the RAA compared to the HBF,
as each sULA is responsible for a narrower angle range.
These results validate that the proposed RAA is a cost-effective method for improving performance, and is highly suitable for edge user communications.

\begin{figure}[htbp]
\centering
	\includegraphics[width=0.55\linewidth]{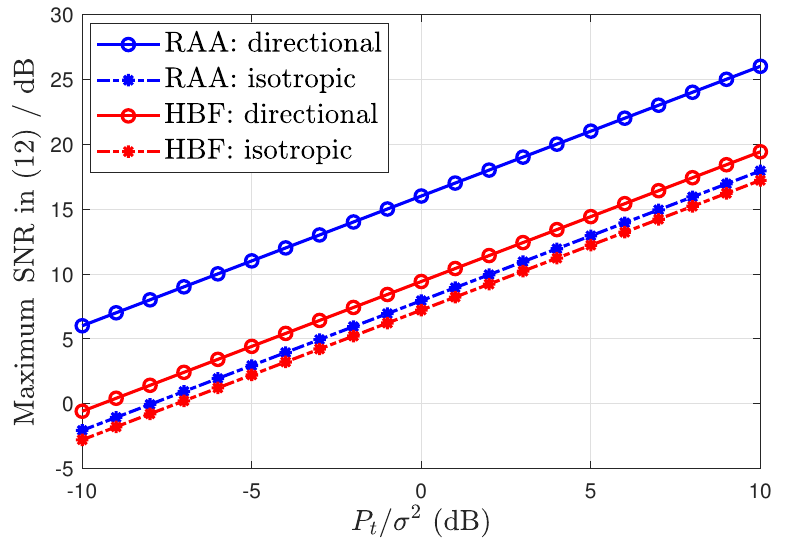}
\caption{Maximum $\text{SNR}$ (dB) in (\ref{SNR2}) for the RAA and HBF.} 
\label{Single}
\end{figure}

Fig. \ref{Single} gives the maximum $\text{SNR}$ (dB) in (\ref{SNR2}) for the proposed RAA and the HBF, considering both isotropic and directional antenna elements.
It can be observed from Fig. \ref{Single} that when using the isotropic antenna, the RAA achieves approximately the same performance as the HBF.
Furthermore, Fig. \ref{Single} also shows that when the directional antenna is used, the SNR of the RAA is around 6 dB higher than that of the HBF at different transmit SNR levels.
This significant performance gain of the RAA is attributed to the higher directional antenna gain.
These results demonstrate that the proposed RAA can enhance the performance of single-user communications compared to the HBF.
\vspace{-0.25cm}
\vspace{-0.25cm}

\begin{figure}[htbp]
\centering
	\includegraphics[width=0.57\linewidth]{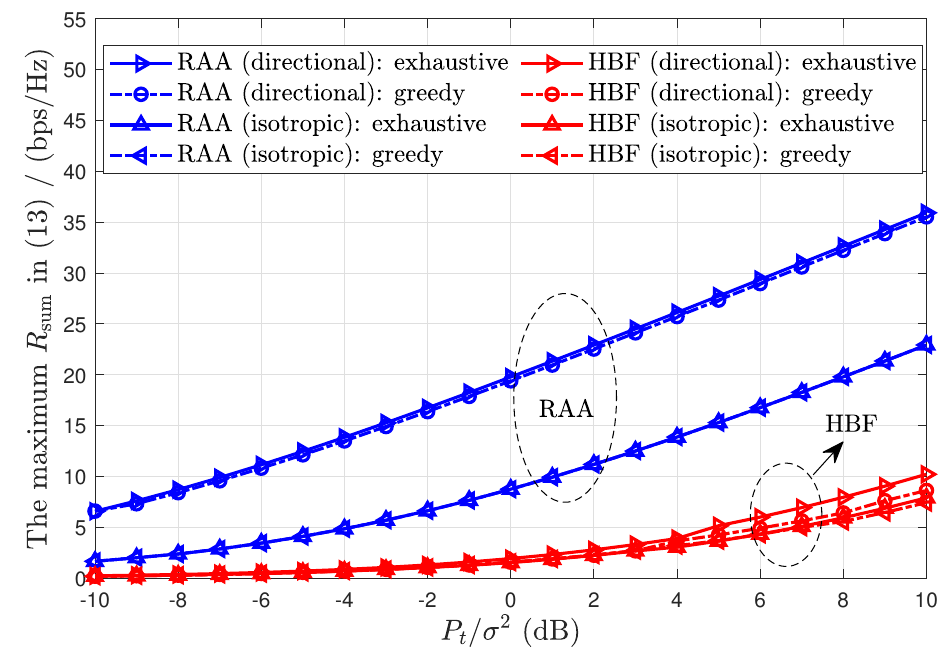}
\caption{The maximum $R_{\text{sum}}$ (bps/Hz) in (\ref{channel model4}) obtained via the exhaustive search and greedy scheme for the RAA and HBF.} 
\label{multiple}
\vspace{-0.4cm}
\end{figure}

Fig. \ref{multiple} illustrates the achievable communication sum rates $R_{\text{sum}}$ (bps/Hz) in (\ref{channel model4}) obtained via exhaustive search and the proposed greedy scheme for the RAA and HBF, considering both the isotropic and directional antenna elements.
It can be observed from Fig. \ref{multiple} that $R_{\text{sum}}$ achieved by the proposed greedy scheme is approximately equal to that obtained via exhaustive search, thereby
validating the effectiveness of the proposed greedy scheme.
In addition, as shown in Fig. \ref{multiple}, the RAA outperforms the HBF in terms of communication performance, particularly when directional antenna elements are employed.
This improvement is attributed to the higher directional antenna gain and the suitability of the RAA's angle-uniform sampling for scenarios where the
users are spatially evenly distributed.
These results demonstrate that the proposed RAA can significantly improve communication performance compared to the HBF. 
\vspace{-0.2cm}

\section{Conclusion}
In this paper, we proposed a novel RAA architecture that enhances wireless communication in a cost-effective manner.
By selecting appropriate sULAs, the system achieves efficient beamforming without relying on analog or digital beamforming.
Compared to the traditional HBF architecture, the RAA significantly reduces hardware costs, improves spatial resolution,  and enhances system performance. 
Simulation results validate that the proposed RAA outperforms the conventional HBF in terms of spatial resolution and communication performance. These findings highlight the potential of the RAA as a practical and scalable solution for future wireless communication and sensing systems.
\vspace{-0.25cm}
\section*{acknowledgment}
This work was supported by the Natural Science Foundation for Distinguished Young Scholars of Jiangsu Province with grant number BK20240070.
\vspace{-0.25cm}

\bibliographystyle{IEEEtran}
\bibliography{./header_short,./bibliography1}

\begin{thebibliography}{10}
\providecommand{\url}[1]{#1}
\csname url@samestyle\endcsname
\providecommand{\newblock}{\relax}
\providecommand{\bibinfo}[2]{#2}
\providecommand{\BIBentrySTDinterwordspacing}{\spaceskip=0pt\relax}
\providecommand{\BIBentryALTinterwordstretchfactor}{4}
\providecommand{\BIBentryALTinterwordspacing}{\spaceskip=\fontdimen2\font plus
\BIBentryALTinterwordstretchfactor\fontdimen3\font minus
  \fontdimen4\font\relax}
\providecommand{\BIBforeignlanguage}[2]{{%
\expandafter\ifx\csname l@#1\endcsname\relax
\typeout{** WARNING: IEEEtran.bst: No hyphenation pattern has been}%
\typeout{** loaded for the language `#1'. Using the pattern for}%
\typeout{** the default language instead.}%
\else
\language=\csname l@#1\endcsname
\fi
#2}}
\providecommand{\BIBdecl}{\relax}
\BIBdecl

\bibitem{xiao2022overview}
Z.~Xiao and Y.~Zeng, ``An overview on integrated localization and communication
  towards $\text{6G}$,'' \emph{Sci. China Inf. Sci.}, vol.~65, no.~3, pp.
  1--46, 2022.

\bibitem{10496996}
H.~Lu, Y.~Zeng, C.~You \emph{et~al.}, ``A tutorial on near-field
  $\text{XL-MIMO}$ communications towards $\text{6G}$,'' \emph{IEEE Commun.
  Surveys Tuts.}, vol.~26, no.~4, pp. 2213--2257, 2024.

\bibitem{ngo2017cell}
H.~Q. Ngo, A.~Ashikhmin, H.~Yang \emph{et~al.}, ``{Cell-free massive MIMO
  versus small cells},'' \emph{IEEE Trans. Wireless Commun.}, vol.~16, no.~3,
  pp. 1834--1850, 2017.

\bibitem{li2024sparse}
X.~Li, H.~Min, Y.~Zeng \emph{et~al.}, ``{Sparse MIMO for ISAC: New
  opportunities and challenges},'' \emph{IEEE Wireless Commun. Mag.}, 2024,
  doi:10.1109/MWC.001.2400201.

\bibitem{zeng2024tutorial}
Y.~Zeng, J.~Chen, J.~Xu \emph{et~al.}, ``{A tutorial on environment-aware
  communications via channel knowledge map for 6G},'' \emph{IEEE Commun.
  Surveys Tuts}, no.~3, pp. 1478--1519, 2024.

\bibitem{wang2009beam}
J.~Wang, Z.~Lan, C.-w. Pyo \emph{et~al.}, ``{Beam codebook based beamforming
  protocol for multi-Gbps millimeter-wave WPAN systems},'' \emph{IEEE J. Sel.
  Areas Commun.}, vol.~27, no.~8, pp. 1390--1399, 2009.

\bibitem{el2014spatially}
O.~El~Ayach, S.~Rajagopal, S.~Abu-Surra \emph{et~al.}, ``{Spatially sparse
  precoding in millimeter wave MIMO systems},'' \emph{IEEE Trans. Wireless
  Commun.}, vol.~13, no.~3, pp. 1499--1513, 2014.

\bibitem{7370753}
R.~Méndez-Rial, C.~Rusu, N.~González-Prelcic \emph{et~al.}, ``{Hybrid MIMO
  architectures for millimeter wave communications: Phase shifters or
  switches?}'' \emph{IEEE Access}, vol.~4, pp. 247--267, 2016.

\bibitem{zeng2016millimeter}
Y.~Zeng and R.~Zhang, ``{Millimeter wave MIMO with lens antenna array: A new
  path division multiplexing paradigm},'' \emph{IEEE Trans. Commun.}, vol.~64,
  no.~4, pp. 1557--1571, 2016.

\bibitem{wong2020fluid}
K.-K. Wong, A.~Shojaeifard, K.-F. Tong \emph{et~al.}, ``Fluid antenna
  systems,'' \emph{IEEE Trans. Wireless Commun.}, vol.~20, no.~3, pp.
  1950--1962, 2020.

\bibitem{zhu2023movable}
L.~Zhu, W.~Ma, and R.~Zhang, ``{Movable antennas for wireless communication:
  Opportunities and challenges},'' \emph{IEEE Commun. Mag.}, vol.~62, no.~6,
  pp. 114--120, 2024.

\bibitem{dong2024movable}
Z.~Dong, Z.~Zhou, Z.~Xiao \emph{et~al.}, ``{Movable antenna for wireless
  communications: Prototyping and experimental results},'' submitted to IEEE
  Trans. Wireless Commun., 2024. Avaliable online:arXiv:2408.08588.

\bibitem{lu2024group}
H.~Lu, Y.~Zeng, S.~Jin \emph{et~al.}, ``Group movable antenna with flexible
  sparsity: Joint array position and sparsity optimization,'' \emph{IEEE
  Wireless Commun. Lett.}, vol.~13, no.~12, pp. 3573--3577, 2024.

\bibitem{pillai2012array}
S.~U. Pillai, \emph{Array signal processing}.\hskip 1em plus 0.5em minus
  0.4em\relax Springer Science \& Business Media, 2012.

\bibitem{qorvo2025}
Qorvo, ``Qorvo official website,'' \url{https://www.qorvo.com/}, 2025,
  accessed: 2025-01-07.

\bibitem{3GPPchannel}
\emph{5G; Study on Channel Model for Frequencies from 0.5 to 100 $\text{GHz}$},
  document 38.901, Version 16.1.0, 3GPP, Technical Specification (TS), Nov.
  2020.

\end{thebibliography}

\end{document}